\documentclass[trackchanges]{aastex7}

\begin{document}
\title{A Long Tidal Feature Extending from a Low-Redshift Galaxy Revealed in Rubin Data Preview 1 Imaging} 

\author[0009-0007-3177-0541]{Virginia Johnson}
\affiliation{Department of Astronomy \& Astrophysics, University of California Santa Cruz, 1156 High Street, Santa Cruz, CA 95064, USA}
\email{vijjohns@ucsc.edu}

\author[0000-0003-2473-0369]{Aaron J.\ Romanowsky}
\affiliation{Department of Physics \& Astronomy, San Jos\'e State University, One Washington Square, San Jose, CA 95192, USA}
\affiliation{Department of Astronomy \& Astrophysics, University of California Santa Cruz, 1156 High Street, Santa Cruz, CA 95064, USA}
\email{aaromanowsky@sjsu.edu}

\begin{abstract}

Using the Vera C.\ Rubin Observatory Data Preview 1 (DP1), we present the detection and photometric analysis of a stellar stream extending along the minor axis of a massive disk-galaxy, LEDA 751050, at a redshift of $z = 0.1$ in the Extended Chandra Deep Field-South (ECDFS). This feature appears to be an analog of the M31 Giant Stellar Stream, spanning around 130 kpc by 15 kpc, with an estimated stellar mass of  $M_\star = 6 \times 10^{8} M_\odot$, and a mean $g$-band surface brightness (SB) of $\mu_g = 28.6$ mag arcsec$^{-2}$. 
This discovery demonstrates Rubin's ability to reveal low-SB substructures and foreshadows the future capabilities when the full Legacy Survey of Space and Time (LSST) data set is actualized.

\end{abstract}

\section{Introduction} 

In hierarchical galaxy formation, the merging of smaller systems into larger ones leaves behind faint tidal features such as streams, shells, and arcs \citep{MartinezDelgado2023}. Many of these halo substructures are expected to lie below current detection thresholds, becoming visible only at the SB limit of $\sim$~30 mag arcsec$^{-2}$ \citep{JohnstonBullock2008}, and highlighting the need for deeper data sets. The Rubin LSST promises an unprecedented combination of sky coverage and depth, expected to reach an SB limit of $\mu \sim$ 30--31 AB mag arcsec$^{-2}$ \citep{Laine2018}, which will allow us to view the imprints of accretion events across a vast sample of galaxies. 

As a preview of this capability, a stellar stream was discovered around the nearby spiral galaxy M61 in Rubin Virgo First-Look imaging, with around five years LSST depth \citep{Romanowsky2025}. 
However, this imaging is not yet fully released for comprehensive science analysis, and here we focus on DP1 from LSSTComCam commissioning observations \citep{comcam,RTN-095}.
Unlike the Virgo observations, DP1  does not have background subtraction optimized for very extended, nearby galaxies, but appears to be good enough for characterizing low-SB features 
$\lesssim 1$~arcmin in diameter
(e.g., an ultra-diffuse galaxy studied by \citealt{RomanowskyUDG}).

\section{Photometric Analysis and Results}

\begin{figure}
    \centering
    \includegraphics[width=0.5\linewidth]{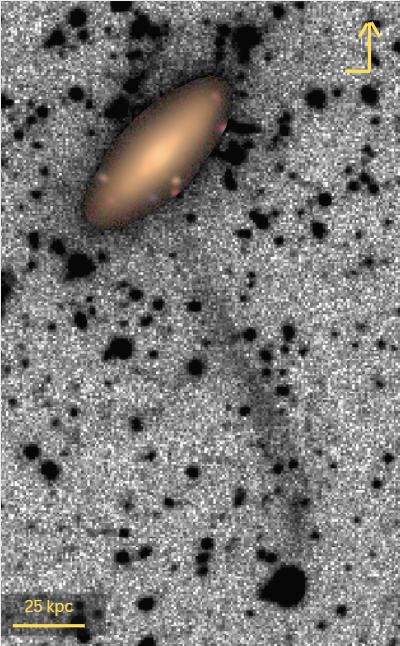}
    \caption{Rubin DP1 negative image of a stellar stream associated with LEDA~751050, using stacked and rebinned $gri$ filters, combined with the color image of the host galaxy from the RSP viewer. A 25-kpc scale-bar is shown.}\label{image}
\end{figure}

We have visually inspected the DP1 $gri$ Hierarchical Progressive Surveys imaging using the Rubin Science Platform's (RSP) Firefly viewer \citep{RSP}, while applying color map stretches, and found a long, polar feature extending from the disk galaxy LEDA~751050 (alternatively PGC~751050; see Figure~\ref{image}).  
This galaxy shows no perturbation from the merger.
It has a spectroscopic redshift of $z=0.102$ from the NASA/IPAC Extragalactic Database,
and lies in the ECDFS, which is the deepest field in DP1.
This field has 211, 230, and 149 visits in $g$, $r$, and $i$ bands, corresponding to total exposure times of 6330, 6900, and 4470 sec, equivalent to 13.2 years of LSST data acquisition \citep{Ivezic2019}. 
We downloaded the coadded $g, r,$ and $ i$ FITS files from the RSP \citep{deepcoadd},
ignoring the $u, z,$ and $y$ bands as they are not deep enough for our analysis.

The apparent tidal stream is centered around the coordinates
03:32:45.6 $-$27:42:53 (J2000), and 
has a length of $\sim 70\arcsec$ ($\sim 130$~kpc) with a width of $\sim 8\arcsec$ ($\sim 15$~kpc).
We sum the stream flux over a large aperture after masking contaminants,
then subtract a background level from averaging over four sky regions,
and adopt a photometric zero-point of 31.4.
We find the stream's mean SB to be $\mu_g = 28.6$, $\mu_r = 28.1$, and $\mu_i = 27.8$~mag~arcsec$^{-2}$, fading at its extrema to detectable limits of $\mu_g \sim 29$~mag~arcsec$^{-2}$.
We then multiplied the SB by the feature's total unmasked area to
obtain apparent magnitudes of $g=21.62$, $r=21.10$, and $i=20.82$,
with uncertainties of $\sim 0.07$~mag mainly from background estimation.
We use the same method with the host galaxy and
find
$g=17.07$, $r= 16.22$, and $i=15.84$ mag. 
The $g$-band flux ratio between the stream and the host,  
after $K$-correction
\citep{2016kcorrcalculator1, 2010kcorrcalculator2},
is therefore 1:77.
The extinction-corrected absolute magnitudes are $M_g = -16.8$ and $-21.6$.
To estimate the stellar mass of the stream, we use the color-based method of
\cite{delosreyes2025},
with $g-r = 0.47$,
and find 
$M_\star = 5.7 \times 10^{8} M_\odot$, which supports a dwarf-galaxy progenitor.
Similarly, we estimate $M_\star = 1.5 \times 10^{11} M_\odot$ for the host galaxy,
leading to a merger mass ratio of $\sim$~1:250. 

The stream in LEDA~751050 is reminiscent of the Giant Southern Stream extending along the minor axis of M31 out to a distance of $\sim$~140~kpc in three dimensions, with a luminosity of $1.5 \times 10^8 L_\odot$ and $\mu_V \sim 30$~mag~arcsec$^{-2}$ \citep{M31original,M31length}. 
Such long streams may be unusual: e.g., in the sample of \cite{Avstreamlength}, none appear to be longer than 50~kpc.
The LEDA~751050 tidal feature represents only a small fraction of what LSST will be capable of, serving as an incredible preview of what may be seen as we approach the survey's full coverage. 

\begin{acknowledgments}
This publication is based in part on proprietary LSST data
\citep{10.71929/rubin/2570308},
and was prepared in accordance with the Rubin Observatory data rights and access policies. All authors of this publication meet the requirements for co-authorship of proprietary LSST data.
This work was supported by the Division of Research and Innovation at San Jos\'e State University (SJSU) under Award Number 25-RSG-08-135. 
The content is solely the responsibility of the authors and does not necessarily represent the official views of SJSU.
\end{acknowledgments}

\facilities{NED, Rubin:Simonyi(LSSTComCam), Rubin:USDAC}

\bibliographystyle{aasjournalv7}
\bibliography{new.ms}{}

\begin{thebibliography}{}
\expandafter\ifx\csname natexlab\endcsname\relax\def\natexlab#1{#1}\fi
\providecommand{\url}[1]{\href{#1}{#1}}
\providecommand{\dodoi}[1]{doi:~\href{http://doi.org/#1}{\nolinkurl{#1}}}
\providecommand{\doeprint}[1]{\href{http://ascl.net/#1}{\nolinkurl{http://ascl.net/#1}}}
\providecommand{\doarXiv}[1]{\href{https://arxiv.org/abs/#1}{\nolinkurl{https://arxiv.org/abs/#1}}}

% type= article
\bibitem[{I.~V. {Chilingarian} {et~al.}(2010){Chilingarian}, {Melchior}, \&
  {Zolotukhin}}]{2016kcorrcalculator1}
{Chilingarian}, I.~V., {Melchior}, A.-L., \& {Zolotukhin}, I.~Y. 2010,
  \bibinfo{title}{{Analytical approximations of K-corrections in optical and
  near-infrared bands},} \mnras, 405, 1409,
  \dodoi{10.1111/j.1365-2966.2010.16506.x}

% type= article
\bibitem[{I.~V. {Chilingarian} \& I.~Y. {Zolotukhin}(2012){Chilingarian} \&
  {Zolotukhin}}]{2010kcorrcalculator2}
{Chilingarian}, I.~V., \& {Zolotukhin}, I.~Y. 2012, \bibinfo{title}{{A
  universal ultraviolet-optical colour-colour-magnitude relation of galaxies},}
  \mnras, 419, 1727, \dodoi{10.1111/j.1365-2966.2011.19837.x}

% type= article
\bibitem[{M.~A.~C. {de los Reyes} {et~al.}(2025){de los Reyes}, {Asali},
  {Wechsler}, {Geha}, {Mao}, {Kado-Fong}, {Pucha}, {Grant}, {Gandhi},
  {Manwadkar}, {Engelhardt}, {Munshi}, \& {Wang}}]{delosreyes2025}
{de los Reyes}, M. A.~C., {Asali}, Y., {Wechsler}, R.~H., {et~al.} 2025,
  \bibinfo{title}{{Stellar Mass Calibrations for Local Low-mass Galaxies},}
  \apj, 989, 91, \dodoi{10.3847/1538-4357/ade4c5}

% type= article
\bibitem[{R. {Ibata} {et~al.}(2001){Ibata}, {Irwin}, {Lewis}, {Ferguson}, \&
  {Tanvir}}]{M31original}
{Ibata}, R., {Irwin}, M., {Lewis}, G., {Ferguson}, A. M.~N., \& {Tanvir}, N.
  2001, \bibinfo{title}{{A giant stream of metal-rich stars in the halo of the
  galaxy M31},} \nat, 412, 49, \dodoi{10.1038/35083506}

% type= article
\bibitem[{{\v{Z}}. {Ivezi{\'c}} {et~al.}(2019){Ivezi{\'c}}, {Kahn}, {Tyson},
  {Abel}, {Acosta}, {Allsman}, {Alonso}, {AlSayyad}, {Anderson}, {Andrew},
  {Angel}, {Angeli}, {Ansari}, {Antilogus}, {Araujo}, {Armstrong}, {Arndt},
  {Astier}, {Aubourg}, {Auza}, {Axelrod}, {Bard}, {Barr}, {Barrau}, {Bartlett},
  {Bauer}, {Bauman}, {Baumont}, {Bechtol}, {Bechtol}, {Becker}, {Becla},
  {Beldica}, {Bellavia}, {Bianco}, {Biswas}, {Blanc}, {Blazek}, {Blandford},
  {Bloom}, {Bogart}, {Bond}, {Booth}, {Borgland}, {Borne}, {Bosch}, {Boutigny},
  {Brackett}, {Bradshaw}, {Brandt}, {Brown}, {Bullock}, {Burchat}, {Burke},
  {Cagnoli}, {Calabrese}, {Callahan}, {Callen}, {Carlin}, {Carlson},
  {Chandrasekharan}, {Charles-Emerson}, {Chesley}, {Cheu}, {Chiang}, {Chiang},
  {Chirino}, {Chow}, {Ciardi}, {Claver}, {Cohen-Tanugi}, {Cockrum}, {Coles},
  {Connolly}, {Cook}, {Cooray}, {Covey}, {Cribbs}, {Cui}, {Cutri}, {Daly},
  {Daniel}, {Daruich}, {Daubard}, {Daues}, {Dawson}, {Delgado}, {Dellapenna},
  {de Peyster}, {de Val-Borro}, {Digel}, {Doherty}, {Dubois},
  {Dubois-Felsmann}, {Durech}, {Economou}, {Eifler}, {Eracleous}, {Emmons},
  {Fausti Neto}, {Ferguson}, {Figueroa}, {Fisher-Levine}, {Focke}, {Foss},
  {Frank}, {Freemon}, {Gangler}, {Gawiser}, {Geary}, {Gee}, {Geha}, {Gessner},
  {Gibson}, {Gilmore}, {Glanzman}, {Glick}, {Goldina}, {Goldstein}, {Goodenow},
  {Graham}, {Gressler}, {Gris}, {Guy}, {Guyonnet}, {Haller}, {Harris},
  {Hascall}, {Haupt}, {Hernandez}, {Herrmann}, {Hileman}, {Hoblitt}, {Hodgson},
  {Hogan}, {Howard}, {Huang}, {Huffer}, {Ingraham}, {Innes}, {Jacoby}, {Jain},
  {Jammes}, {Jee}, {Jenness}, {Jernigan}, {Jevremovi{\'c}}, {Johns}, {Johnson},
  {Johnson}, {Jones}, {Juramy-Gilles}, {Juri{\'c}}, {Kalirai}, {Kallivayalil},
  {Kalmbach}, {Kantor}, {Karst}, {Kasliwal}, {Kelly}, {Kessler}, {Kinnison},
  {Kirkby}, {Knox}, {Kotov}, {Krabbendam}, {Krughoff}, {Kub{\'a}nek},
  {Kuczewski}, {Kulkarni}, {Ku}, {Kurita}, {Lage}, {Lambert}, {Lange},
  {Langton}, {Le Guillou}, {Levine}, {Liang}, {Lim}, {Lintott}, {Long},
  {Lopez}, {Lotz}, {Lupton}, {Lust}, {MacArthur}, {Mahabal}, {Mandelbaum},
  {Markiewicz}, {Marsh}, {Marshall}, {Marshall}, {May}, {McKercher}, {McQueen},
  {Meyers}, {Migliore}, {Miller}, \& {Mills}}]{Ivezic2019}
{Ivezi{\'c}}, {\v{Z}}., {Kahn}, S.~M., {Tyson}, J.~A., {et~al.} 2019,
  \bibinfo{title}{{LSST: From Science Drivers to Reference Design and
  Anticipated Data Products},} \apj, 873, 111, \dodoi{10.3847/1538-4357/ab042c}

% type= article
\bibitem[{K.~V. {Johnston} {et~al.}(2008){Johnston}, {Bullock}, {Sharma},
  {Font}, {Robertson}, \& {Leitner}}]{JohnstonBullock2008}
{Johnston}, K.~V., {Bullock}, J.~S., {Sharma}, S., {et~al.} 2008,
  \bibinfo{title}{{Tracing Galaxy Formation with Stellar Halos. II. Relating
  Substructure in Phase and Abundance Space to Accretion Histories},} \apj,
  689, 936, \dodoi{10.1086/592228}

% type= article
\bibitem[{S. {Laine} {et~al.}(2018){Laine}, {Martinez-Delgado}, {Trujillo},
  {Duc}, {Grillmair}, {Frenk}, {Hendel}, {Johnston}, {Mihos}, {Moustakas},
  {Beaton}, {Romanowsky}, {Greco}, \& {Erkal}}]{Laine2018}
{Laine}, S., {Martinez-Delgado}, D., {Trujillo}, I., {et~al.} 2018,
  \bibinfo{title}{{LSST Cadence Optimization White Paper in Support of
  Observations of Unresolved Tidal Stellar Streams in Galaxies beyond the Local
  Group},} arXiv e-prints, arXiv:1812.04897, \dodoi{10.48550/arXiv.1812.04897}

% type= article
\bibitem[{D. {Mart{\'\i}nez-Delgado} {et~al.}(2023){Mart{\'\i}nez-Delgado},
  {Cooper}, {Rom{\'a}n}, {Pillepich}, {Erkal}, {Pearson}, {Moustakas},
  {Laporte}, {Laine}, {Akhlaghi}, {Lang}, {Makarov}, {Borlaff}, {Donatiello},
  {Pearson}, {Mir{\'o}-Carretero}, {Cuillandre}, {Dom{\'\i}nguez},
  {Roca-F{\`a}brega}, {Frenk}, {Schmidt}, {G{\'o}mez-Flechoso}, {Guzman},
  {Libeskind}, {Dey}, {Weaver}, {Schlegel}, {Myers}, \&
  {Valdes}}]{MartinezDelgado2023}
{Mart{\'\i}nez-Delgado}, D., {Cooper}, A.~P., {Rom{\'a}n}, J., {et~al.} 2023,
  \bibinfo{title}{{Hidden depths in the local Universe: The Stellar Stream
  Legacy Survey},} \aap, 671, A141, \dodoi{10.1051/0004-6361/202245011}

% type= article
\bibitem[{A.~W. {McConnachie} {et~al.}(2003){McConnachie}, {Irwin}, {Ibata},
  {Ferguson}, {Lewis}, \& {Tanvir}}]{M31length}
{McConnachie}, A.~W., {Irwin}, M.~J., {Ibata}, R.~A., {et~al.} 2003,
  \bibinfo{title}{{The three-dimensional structure of the giant stellar stream
  in Andromeda},} \mnras, 343, 1335, \dodoi{10.1046/j.1365-8711.2003.06785.x}

% type= article
\bibitem[{J. {Mir{\'o}-Carretero} {et~al.}(2024){Mir{\'o}-Carretero},
  {Mart{\'\i}nez-Delgado}, {G{\'o}mez-Flechoso}, {Cooper}, {Akhlaghi},
  {Donatiello}, {Kuijken}, {Lang}, {Makarov}, {Laine}, \&
  {Roca-F{\`a}brega}}]{Avstreamlength}
{Mir{\'o}-Carretero}, J., {Mart{\'\i}nez-Delgado}, D., {G{\'o}mez-Flechoso},
  M.~A., {et~al.} 2024, \bibinfo{title}{{Extragalactic stellar tidal streams in
  the Dark Energy Survey},} \aap, 691, A196,
  \dodoi{10.1051/0004-6361/202451685}

% type= techreport
\bibitem[{{NSF-DOE Vera C.\ Rubin Observatory}(2025{\natexlab{a}}){NSF-DOE Vera
  C.\ Rubin Observatory}}]{RTN-095}
{NSF-DOE Vera C. Rubin Observatory}. 2025{\natexlab{a}}, {The Vera C. Rubin
  Observatory Data Preview 1}, {Rubin Technical Note} RTN-095, {NSF-DOE Vera C.
  Rubin Observatory}, \dodoi{10.71929/rubin/2570536}

% type= misc
\bibitem[{{NSF-DOE Vera C.\ Rubin Observatory}(2025{\natexlab{b}}){NSF-DOE Vera
  C.\ Rubin Observatory}}]{deepcoadd}
{NSF-DOE Vera C. Rubin Observatory}. 2025{\natexlab{b}}, {Legacy Survey of
  Space and Time Data Preview 1: deep\_coadd dataset type [Data set]}, NSF-DOE
  Vera C. Rubin Observatory, \dodoi{10.71929/RUBIN/2570313}

% type= misc
\bibitem[{{NSF-DOE Vera C.\ Rubin Observatory}(2025{\natexlab{c}}){NSF-DOE Vera C.\ Rubin
  Observatory}}]{10.71929/rubin/2570308}
{NSF-DOE Vera C.\ Rubin Observatory}. 2025{\natexlab{c}}, {Legacy Survey of Space and Time
  Data Preview 1 [Data set]}, NSF-DOE Vera C. Rubin Observatory,
  \dodoi{10.71929/RUBIN/2570308}

% type= article
\bibitem[{A.~J. Romanowsky {et~al.}(2025{\natexlab{b}})Romanowsky, Martínez-Delgado,
  Donatiello, Miró-Carretero, \& Laine}]{Romanowsky2025}
Romanowsky, A.~J., Martínez-Delgado, D., Donatiello, G., Miró-Carretero, J.,
  \& Laine, S. 2025{\natexlab{b}}, \bibinfo{title}{A Stellar Stream around the Spiral Galaxy
  Messier 61 in Rubin First Look Imaging,} Research Notes of the AAS, 9, 292,
  \dodoi{10.3847/2515-5172/ae1829}

% type= article
\bibitem[{A.~J. {Romanowsky} {et~al.}(2025{\natexlab{a}}){Romanowsky}, {Tang}, \&
  {Bundy}}]{RomanowskyUDG}
{Romanowsky}, A.~J., {Tang}, Y., \& {Bundy}, K.~A. 2025{\natexlab{a}},
  \bibinfo{title}{{Morphology and Stellar Populations of a Candidate
  Ultra-diffuse Galaxy in Early Euclid and Rubin Imaging},} Research Notes of
  the American Astronomical Society, 9, 181, \dodoi{10.3847/2515-5172/adee10}

% type= misc
\bibitem[{{SLAC National Accelerator Laboratory} \&  {NSF-DOE Vera C.\ Rubin
  Observatory}(2024){SLAC National Accelerator Laboratory} \& {NSF-DOE Vera C.\
  Rubin Observatory}}]{comcam}
{SLAC National Accelerator Laboratory}, \& {NSF-DOE Vera C. Rubin Observatory}.
  2024, LSST Commissioning Camera, SLAC National Accelerator Laboratory (SLAC),
  Menlo Park, CA (United States), \dodoi{10.71929/RUBIN/2561361}

% type= techreport
\bibitem[{{Vera C. Rubin Observatory Science Pipelines Developers}(2025){Vera
  C.\ Rubin Observatory Science Pipelines Developers}}]{RSP}
{Vera C.\ Rubin Observatory Science Pipelines Developers}. 2025, {The LSST
  Science Pipelines Software: Optical Survey Pipeline Reduction and Analysis
  Environment}, {Project Science Technical Note} PSTN-019, {Vera C. Rubin
  Observatory}, \dodoi{10.71929/rubin/2570545}

\end{thebibliography}

\end{document}